\documentclass[aip,apl,reprint]{revtex4-1}

\usepackage{graphicx}
\usepackage{dcolumn}%
\usepackage{bm}
\usepackage{amsfonts}
\usepackage{amsfonts}
\usepackage{amssymb}
\usepackage{amsmath}
\usepackage{color}
\usepackage{epsfig}
\usepackage[normalem]{ulem}

\newcommand{\e}{\varepsilon}
\newcommand{\w}{\omega}
\newcommand{\im}{\mathrm{i}}
\newcommand{\Einc}{\textbf{E}_{\text{inc}}}
\newcommand{\Hinc}{\textbf{H}_{\text{inc}}}
\newcommand{\Etra}{\textbf{E}_{\text{trans}}}
\newcommand{\Htra}{\textbf{H}_{\text{trans}}}

\begin{document}

\title{Manipulating polarized light with a planar slab of Black Phosphorus}

\author{Constantinos A. Valagiannopoulos}
\email[]{konstantinos.valagiannopoulos@nu.edu.kz}
\homepage[]{http://sst.nu.edu.kz/konstantinos-valagiannopoulos}
\affiliation{Department of Physics, Nazarbayev University, Astana, KZ-010000, Kazakhstan}
\author{Marios Mattheakis}
\email[]{mariosmat@g.harvard.edu \hspace{3cm}}
\homepage[]{http://scholar.harvard.edu/marios$\_$matthaiakis}
\affiliation{School of Engineering and  Applied Sciences, Harvard University, Cambridge, MA-02138, USA}
\affiliation{Department of Physics, University of Crete, Heraklion, GR-71003, Greece}
\author{ Sharmila N. Shirodkar}
\affiliation{Department of Materials Science \& Nano-Engineering, Rice University, Houston, TX-77005, USA}
\author{ Efthimios Kaxiras}
\affiliation{School of Engineering and  Applied Sciences, Harvard University, Cambridge, MA-02138, USA}
\affiliation{Department of Physics, Harvard University, Cambridge, MA-02138, USA}

\begin{abstract}
Wave polarization contains valuable information for electromagnetic signal processing; thus, the ability to manipulate it, can be extremely useful in modeling photonic devices. In this work, we propose designs comprised of one of the emerging and interesting two-dimensional media: Black Phosphorus. Due to substantial in-plane anisotropy, a single slab of Black Phosphorus can be very efficient for controlling the polarization state of electromagnetic waves. We investigate Black Phosphorus slabs that filter the fields along one direction, or achieve {polarization axis} rotation, or convert linear polarization to circular. These slabs can be employed as components in numerous mid-IR integrated structures.  \\
\end{abstract}

\maketitle

The polarization of electromagnetic plane waves is a feature which contains valuable information for the carried signal and thus directly pertains to the operation and performance of  electronic and photonic devices. Since any field distribution can be expressed as Fourier spatial integrals of plane waves, the direction of electric and magnetic field as a function of time is a crucial characteristic regardless of the source. Most electromagnetic structures and applications are polarization-sensitive, including directional couplers of Surface Plasmon Polaritons \cite{PolConSPP}, beam steering gratings \cite{PolGratings}, optical controlling of magnetic media \cite{FerroNano}, extreme events in optical disorder systems \cite{mariosRWs} and bent radiation from gradient index lens networks \cite{mariosLLW} or plasmonic nanoantennas \cite{LightBending}. Furthermore, several electrodynamic devices work only for waves of specific polarization type such as asymmetric transmitters of linearly polarized light \cite{TransmissionPol}, fully controllable composite metascreens \cite{ControlMetascreen}, all-angle ultra-violet lenses operating with circular polarization \cite{FlatLensing} and non-linear semiconducting heterostructures \cite{NonlinearResponse}.

Given the importance of the polarization of light, it is worthwhile to explore devices which can convert it from one type to another (linear, circular, elliptical). During the last decade, several designs  have been presented covering the whole operational frequency range from broadband THz converters \cite{LinPolConversionTHz} and chiral structures achieving diffraction-based polarization tilts \cite{PolChiral} to microwave polarization rotators \cite{PolarizationRotator}. The most common way to fabricate such designs is to use the metamaterial paradigm: a huge number of electrically tiny particles are distributed, usually homogeneously, into a volume in order to create the desired electromagnetic regime and the necessary directional selectivity to affect the polarization of electromagnetic waves \cite{PolAnisMetam}. In the same spirit, one can construct metasurfaces by distributing the unit cells on a surface boundary (instead of a volume domain) providing multi-beam reflections with independent control of polarizations \cite{PolAnisoMetasurfaces}, enhanced asymmetric transmission \cite{PolTransSRR}, or conversion of circularly polarized light into its cross-polarized counterpart \cite{PlanarPhotonics}.

Instead of employing artificial structures, like the aforementioned ones, it would be very interesting to use actual media which can perform similar polarization rotations and manipulations. The most suitable materials that can serve such a purpose are the two-dimensional (2D) media which offer unique possibilities both in terms of model designing and prototype fabrication. Beginning from graphene \cite{RiseOfGraphene} which revealed fascinating properties useful in multiple applications, other 2D media such as Molybdenium Sulfide (MoS$_2$) \cite{CvsMoS2} can be combined with graphene to form integrated wave-steering components \cite{MyJ74}. Hexagonal Boron Nitride (hBN) is also used as a key 2D medium \cite{RamanEnh} together with other crystals \cite{Stacking2DMedia, BeyondGraphene} to properly tailor the monolayers in the quest of unprecedented electromagnetic properties. As far as manipulating the polarization of waves is concerned, which constitutes the primary aim of this study, Black Phosphorus (BP) is the most befitting two-dimensional medium due to its substantial in-plane anisotropy \cite{SemicondingBP, ZhangNanoTech, YePhosphorine}. BP has been extensively studied in terms of its electrostatic screening properties \cite{PlasmonsBP} and its response to external strain of the epitaxial lattice \cite{StrainInducedBP}.

In this work, we use a stack of BP monolayers, forming a planar finite-thickness BP film, which is illuminated by a normally incident plane wave of arbitrary linear polarization. We confine our research to mid-infrared (IR) frequencies ($\w=$ 100$\pi$--240$\pi$ THz) for which we determine the dielectric function of BP from  ab-initio calculations. Based on this information, we notice  the contrast between the permittivities along the two directions parallel to the monolayers. This allows us to determine the thickness of the film slab at specific operational frequencies so that efficient filtering of one of the two electric components is achieved. We also obtain the optimal configurations,  concerning the tilt of the polarization direction and the conversion of the incident linear polarization to quasi-circular (elliptical with small eccentricity). The proposed designs form the basis for experimental efforts in fabricating polarization-based signal processing components for terahertz applications.

BP comprises atoms of Phosphorus (P) located in the three-dimensional space so that an extremely thin though metastable layer can be exfoliated and exist as a two-dimensional structure. Once this lattice of electrically neutral atoms is placed into a background electric field $\textbf{E}$, the atoms are polarized, and by exerting forces and torques to each other, twist the atomic lattice. This macroscopic displacement of charges creates a polarization field, which is proportional to the background one \cite{Orfanidis}, written as: $\left([\e]-\textbf{I}\right)\cdot\textbf{E}$, where $\textbf{I}$ is the $3\times 3$ identity matrix and  $[\e]$ is the relative effective permittivity tensor. In the case of bulk BP, $[\e]$ takes the diagonal form $[\e]={\rm diag}(\e_x, \e_y, \e_z)$ containing the axial permittivities. The ab-initio estimation of the complex parameters $(\e_x, \e_y, \e_z)$ for a field time-dependence of the form $\exp(-i \w t)$, is made by suitably evaluating the interaction between ionic cores and valence electrons in minimal-energy stable configurations with specific locations for the P atoms. Weak van der Waals interactions permit 2D materials, such as BP, to form bulk media by stacking multiple layers on top of each other, with optical properties \cite{MyJ74, rodrick, kaxirasMoS2} substantially different from the behavior of a  single layer.

We carried out the first-principles density functional (DFT) calculations  using the Quantum ESPRESSO package \cite{Sharmila1}, with plane-wave basis sets and norm-conserving pseudopotentials  \cite{Sharmila2} representing the interaction between the ionic cores and valence electrons. The exchange-correlation energy of the electrons is treated within the Generalized Gradient Approximated (GGA) functional of Becke-Lee-Yang-Parr (BLYP) \cite{Sharmila3, Sharmila4}. For the plane-wave basis we use an energy cutoff of 70 Ry for the wavefunctions and 280 Ry for the charge density. Structures were determined through minimization of the energy until the Hellmann-Feynman forces on each atom were smaller in magnitude than 0.03 eV/{\AA}. The semi-empirical Grimme's DFT-D2 functional \cite{Sharmila5} was employed to include the van der Waals interactions between the layers. The Brillouin zone integrations were carried over $12\times 9\times 4$ Monkhorst-Pack \cite{Sharmila6} set of k-points for structural relaxation, while a denser $20\times 15\times 6$ uniform set of k-points was used in calculating the dielectric function.

The lattice constants a, b, c {(along $x,y,z$ axes correspondingly)} of bulk BP after relaxation were found to be 3.36 {\AA}, 4.68 {\AA} and 11.21 {\AA}, respectively. The corresponding errors in comparison with experimental values are 1.5$\%$, 7.0$\%$ and 7.0$\%$ higher than usual DFT calculations, since GGA functionals do not accurately describe the electronic properties and the structure of bulk BP \cite{Sharmila7}. Other functionals, such as Heyd-Scuseria-Ernzerhof (HSE06), are very computationally expensive for the determination of dielectric function, so we resorted to using the atomic and electronic structure determined utilizing the BLYP functional which at least gives a finite band gap.

\begin{figure}[ht]
\centering
\includegraphics[scale=0.26]{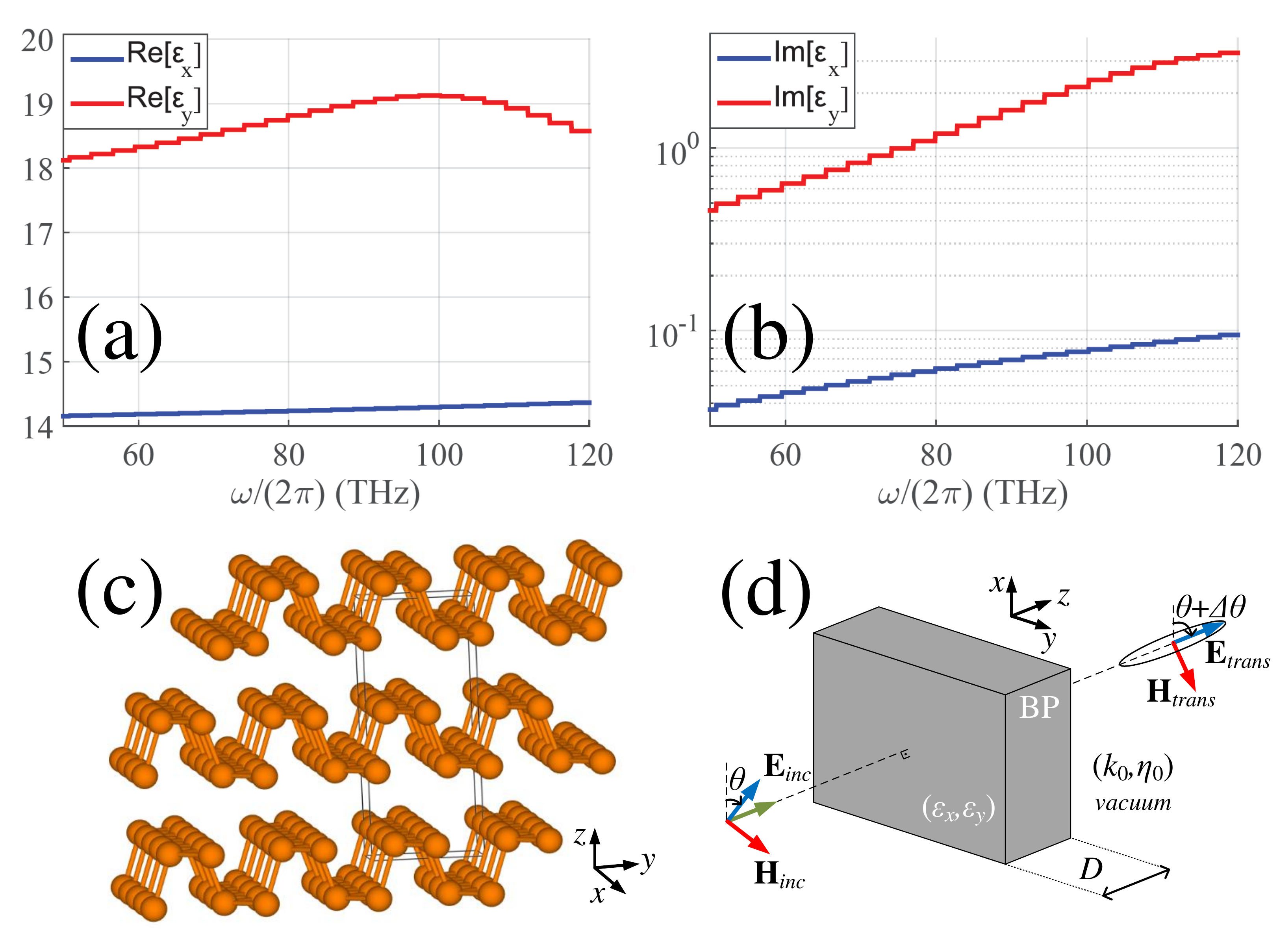}
\caption{{The: (a) real and (b) imaginary part of the complex dielectric function of bulk BP for a continuous frequency spectrum $50~{\rm THz}<\w/(2\pi)<120~{\rm THz}$. Ab-initio values are illustrated for $\e_x$ (blue) and $\e_y$ (red). (c) The BP atomic crystal together with the employed Cartesian coordinate system $(x,y,z)$ and an orthorhombic  unit cell of BP. (d) A BP slab of thickness $D$ is built by stacking of BP monolayers. The slab is  excited by a normally incident plane wave with arbitrary polarization angle $\theta$. ($\Einc, \Hinc$) stand for the incident electromagnetic field components and ($\Etra, \Htra$) the transmitted field. The reflected field is not shown for brevity.}}
\label{fig:Figs1}
\end{figure}

{In Figs. \ref{fig:Figs1}(a) and \ref{fig:Figs1}(b), we show the variations of the real and imaginary parts of the in-plane (transversal) relative dielectric function (directions on the $xy$ plane) with respect to ${\rm THz}$ frequencies into the range $50~{\rm THz}<\w/(2\pi)<120~{\rm THz}$. The substantial in-plane anisotropy of BP layers is evident since both real and imaginary parts of $(\e_x,\e_y)$ differ from each other. Most importantly, $\e_x$ is almost lossless compared to $\e_y$ as represented in the logarithmic scale of Fig. \ref{fig:Figs1}(b). The step discontinuities in the curves are due to the finite number of ab-initio calculations (numerical meshing) and do not affect the validity of the presented results. In Fig. \ref{fig:Figs1}(c), we depict a schema for the BP crystal and notice that the low-loss behavior ($x$ axis) occurs along the so-called ``zigzag'' direction, contrary to the very lossy one ($y$ axis) appearing along the ``armchair'' direction of the atomic lattice. In the same Fig. \ref{fig:Figs1}(c), one can observe the orthorhombic unit cell of BP with dimensions a, b and c as defined above.} 

{ The used permittivities $(\e_x,\e_y)$ are not taken dependent on the thickness $D$ because  for slabs thicker than some nanometers (a few dozens of monolayers), the dielectric function converges to a specific value (bulk BP). The same happens for other 2D materials like MoS$_2$ and hBN \cite{rodrick, Andersen}. Since in the present work we focus on effects occurred for large $D>1000$ nm, namely to more than 2000 BP layers (the thickness of a BP monolayer is about 5 {\AA}), there is no substantial need for employing thickness-dependent BP permittivity. Our static dielectric constants are in agreement with experimental results\cite{Nagahama}. Furthermore, theoretical formulations \cite{Asahina} verify the relationship between different directions shown in Figs. \ref{fig:Figs1}(a) and \ref{fig:Figs1}(b) even though they underestimate the experimental static dielectric constant.} 

We consider the simplest possible configuration comprised by a rectangular volume of BP, fabricated by stacking many single layers  one upon the other to acquire a finite thickness $D$ as shown in Fig. \ref{fig:Figs1}(d). To examine what kind of wave manipulations are possible by such a free standing BP film, due to the in-plane effective anisotropy of this medium, we excite it by an electromagnetic wave having both $x$ and $y$ electric components.  In particular, the incident electric field in vacuum having wavenumber $k_0=\w/c$ is linearly polarized, and propagates normally to the air-BP interface. Hence, it is written in the form: $\Einc=E_0\left(\cos\theta~\hat{\textbf{x}}+\sin\theta~\hat{\textbf{y}}\right)e^{\im k_0z}$ where $E_0>0$ is the amplitude measured in {\rm V/m}. The transmitted wave propagates into vacuum from the other side of the BP layer, with an electric field given by: $\Etra=E_0\left(T_x\hat{\textbf{x}}+T_y\hat{\textbf{y}}\right)e^{\im k_0z}$. The transmission coefficients $(T_x,T_y)$ are found after applying the boundary conditions across the front ($z=-D$) and rear ($z=0$) interfaces between vacuum and BP, as follows: 
\begin{equation}
\label{eq:TransmissionCoefficientX}
T_x=\frac{4~e^{-\im k_0D}\sqrt{\e_x}\cos\theta}
{e^{-\im k_0 D \sqrt{\e_x}}\left(\sqrt{\e_x}+1\right)^2-e^{\im k_0D \sqrt{\e_x}}\left(\sqrt{\e_x}-1\right)^2}, 
\end{equation}
\begin{equation}
\label{eq:TransmissionCoefficientY}
T_y=\frac{4~e^{-\im k_0D}\sqrt{\e_y}\sin\theta}
{e^{-\im k_0D\sqrt{\e_y}}\left(\sqrt{\e_y}+1\right)^2-e^{\im k_0D\sqrt{\e_y}}\left(\sqrt{\e_y}-1\right)^2}.
\end{equation}
The complex numbers $(T_x,T_y)$  determine the characteristics of the output (transmitted) beam.  A wave with such amplitudes would be elliptically polarized of ellipse eccentricity $e_T$, which is tilted \cite{Orfanidis} by $\Delta\theta$ from the initial direction of the incidence angle $\theta$. Expressing the transmission coefficients in polar form, $T_s=|T_s|e^{\im \phi_s}$ with $s=x$ or $y$, the eccentricity of the elliptical polarization of the transmission reads \cite{Orfanidis}: 
\begin{equation}
e_T=\sqrt{\frac{2g\left(T_x,T_y,\Delta\phi\right) } {\left(|T_x|^2+|T_y|^2\right)^2+g\left(T_x,T_y,\Delta\phi\right) } },
\label{eq:EllipseEccentricity}
\end{equation}
where $g$ is an auxiliary function given by:
\begin{equation}
g(T_x,T_y,\Delta\phi)=|T_x T_y| \sqrt{\frac{|T_x|^2}{|T_y|^2}+\frac{|T_y|^2}{|T_x|^2}+2\cos(2\Delta\phi)}
\end{equation} 
and  $\Delta\phi=\phi_x-\phi_y$.  $|\Delta\theta|$  is computed numerically  to avoid ambiguities between supplementary arcs.

In the following, we explore designs of the BP slab  which manipulate in an interesting way the incoming radiation to produce a transmitted ray characterized by three parameters: $(P_T,e_T,|\Delta\theta|)$, where $P_T=|T_x|^2+|T_y|^2$ is the carried power (as a fraction of the incident power) per m$^2$. We consider the frequency range $50~{\rm THz}<\w/(2\pi)<120~{\rm THz}$, and a physical length for the slab that is { comprised of numerous BP monolayers but simultaneously} kept below a free-space wavelength: $1~{\rm \mu m}<D<3~{\rm \mu m}$, and examine the EM propagation  under every possible linear polarization angle $0<\theta<180^o$. { Our analysis concerns much higher frequencies than similar works \cite{TonyLow}, where the reported effects are demonstrated in the vicinity of 10 THz.}

\begin{figure}[ht]
\centering
\includegraphics[scale=0.26]{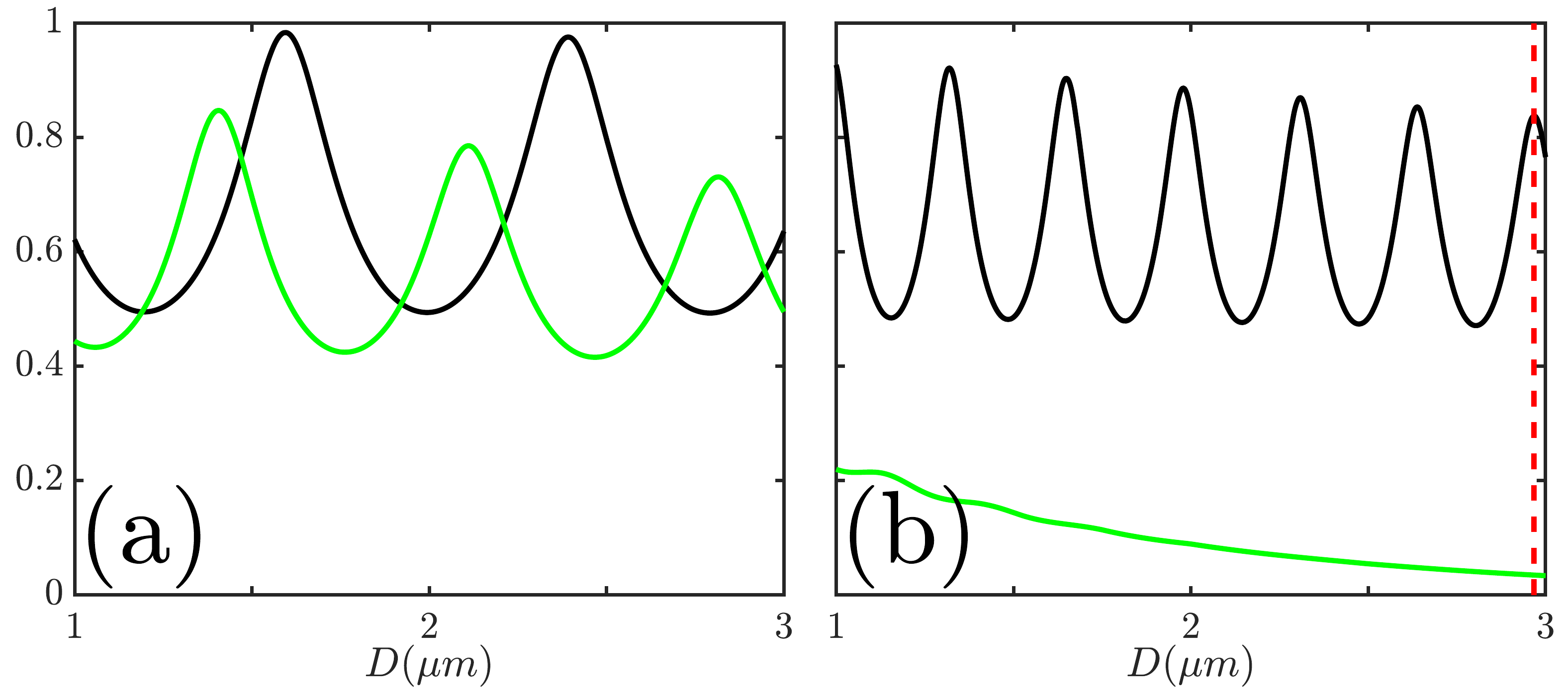}
\caption{The variation of the magnitude of the reduced transmission coefficients $|T_x|/|\cos\theta|$ (black) and $|T_y|/|\sin\theta|$ (green) with respect to the thickness $D$ of the BP slab for: (a) $\omega=100\pi$ THz, (b) $\omega=240\pi$ THz.}
\label{fig:Figs2}
\end{figure}

Figs. \ref{fig:Figs1}(a) and \ref{fig:Figs1}(b) shows the large contrast between the permittivity losses along the two in-plane axes of bulk BP. This sizeable difference reveals a potential use of BP layers as a filter for one of the two electrical components: the one parallel to the $y$ axis. The variation of the transmission magnitude $|T_x/\cos\theta|$ and $|T_y/\sin\theta|$, as function of the slab thickness, for the lowest ($\omega=100\pi$ THz) and highest ($\omega=240\pi$ THz) frequencies used in Figs. \ref{fig:Figs1}(a) and \ref{fig:Figs1}(b), are shown in Fig. \ref{fig:Figs2}(a) and  \ref{fig:Figs2}(b), respectively. These quantities do not depend on polarization angle $\theta$ as indicated by (\ref{eq:TransmissionCoefficientX}), (\ref{eq:TransmissionCoefficientY}). The magnitude of the transmission coefficient $|T_x/\cos\theta|$ which accounts for the direction with small losses of the corresponding permittivity $\e_x$, is oscillating with respect to $D$; this behavior is referred to as Fabry-Perot resonance \cite{FabryPerot12}. 

This is not the case for the transmission coefficient of the other direction (electric field parallel to $y$ axis), especially for higher frequencies GIVING more significant losses. For higher frequencies (Fig. \ref{fig:Figs2}(b)), where ${\rm Im}[\e_y]$ is larger, the quantity $|T_y/\sin\theta|$ gets attenuated and this damping effect is more substantial for thicker samples $D$. The largest used frequencies (namely close to $\w=240\pi$ THz) result in the highest losses ${\rm Im}[\e_y]$ and selecting the right value of $D$ can lead to a maximum $|T_x|$. This value is indicated with a dashed red line in Fig. \ref{fig:Figs2}(b). For such a configuration, a BP slab is expected to filter out the $y$ electric component of the incident radiation.

\begin{figure}[ht]
\centering
\includegraphics[scale =0.32]{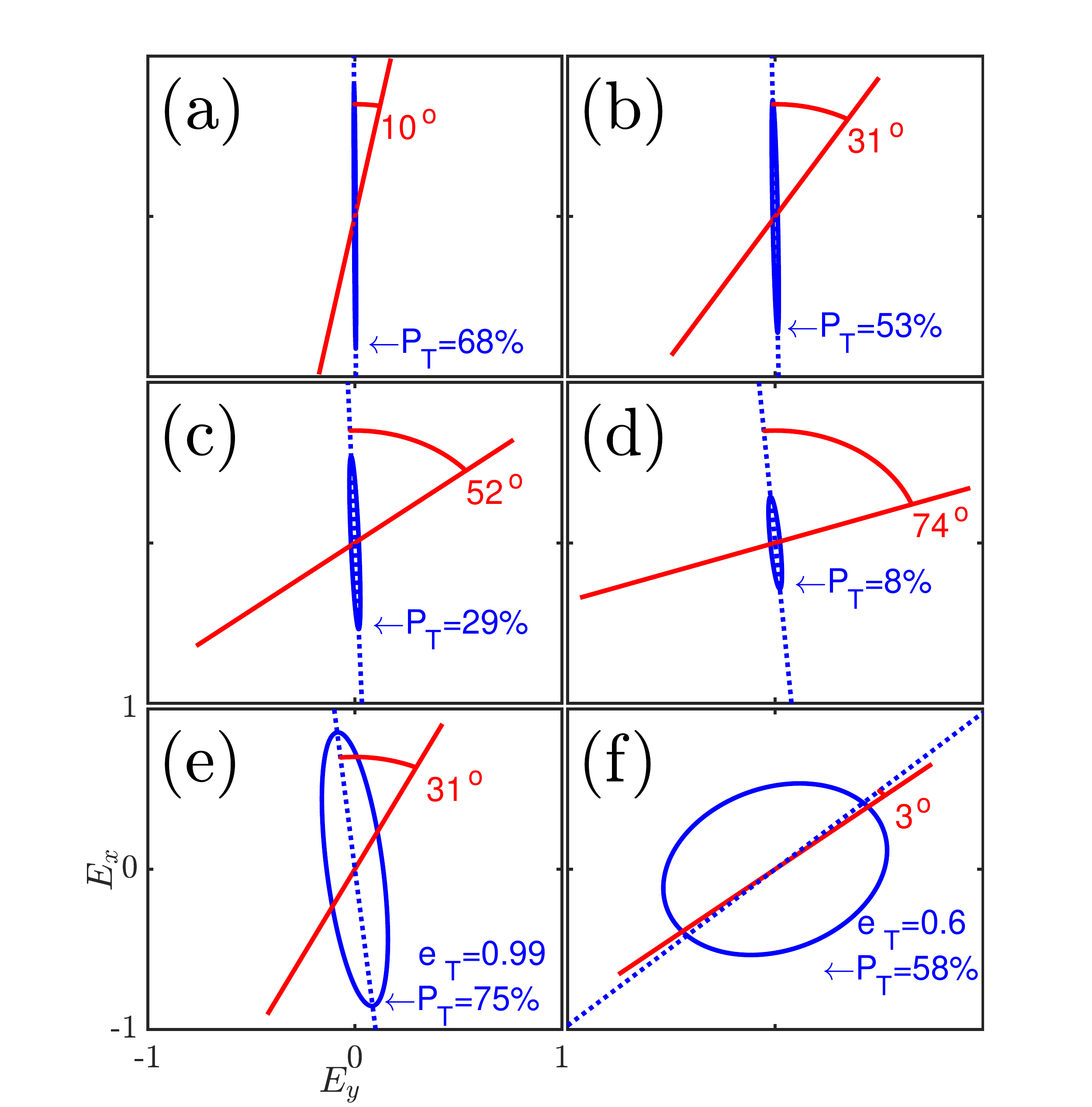}
\caption{The patterns sketched by the tips of the electric field vectors, $\Einc$ (red), $\Etra$ (blue). The output power is given by $P_T$ as a fraction of input power, the eccentricity of the output field by is given by $e_T$, and the tilting  angle is shown {in degrees}. Results are represented  for various incidence polarization angles $\theta$: (a) $\theta=10^o$, (b) $\theta=30^o$, (c) $\theta=50^o$ and (d) $\theta=70^o$. Plot parameters: $\w=240\pi$ THz, $D$=2968 nm. (e) Large {polarization axis} rotation for $\theta=25^o$, $\w=148\pi$ THz and $D=2685$ nm.  (f)  Conversion of linear to quasi-circular polarization for $\theta=49^o$, $\w=100\pi$ THz and $D=1492$ nm.}
\label{fig:Figs3}
\end{figure}

In  Figs. \ref{fig:Figs3} with the observer located at a fixed $(x,y)$ intersecting plane and normal to the $z$ direction of propagation, we show the shape described by the tip of the electric field vectors, incident  $\Einc$ (red) and transmitted $\Etra$ (blue), as well as the major polarization direction of the output field (dashed line). In Fig. \ref{fig:Figs3}(a), where we consider an incidence polarization angle $\theta=10^o$, the output signal is $x$-polarized and the magnitude of the component $T_x$ is very close to the corresponding projection ($\cos\theta$) of its incident counterpart. In Fig. \ref{fig:Figs3}(b), where we use plane-wave excitation with $\theta=30^o$, the $y$ electric component of the incident field gets substantially suppressed. When larger polarization angles $\theta$ are considered in Fig. \ref{fig:Figs3}(c) and \ref{fig:Figs3}(d), the transmission field becomes progressively more elliptically polarized, describing ellipses of smaller eccentricity. This is attributed to the fact that the (preserved) $x$ component of $\Einc$ becomes smaller and thus comparable with the remainder of the $y$ component which survives the filtering; thus, $|T_y|$ is no longer negligible compared to $|T_x|$ and $e_T$ diverges from unity. Note that the cancellation of the $y$ component is imperfect  since $|T_y|$ cannot be zero through a lossy homogeneous medium. Similarly, for the $x$ component we have $|T_x|<\cos\theta$, meaning that a small part of the $x$-directed field power is converted into thermal form due to the low but nonzero ${\rm Im}[\e_x]$.   

Another potential operation of the BP slab is tilting of the polarization direction of the incident field by a desired angle. This effect, can be characterized as non-magnetic Faraday rotation \cite{GiantFaradayRotation} and is used in designing isolators which prevent the multiple propagation of signals, as well as modulators, where the information is hidden in the tilt polarization angle of the output. {Note that the term ``Faraday rotation'' is used mainly for structures with magnetic bias which gives non-reciprocal properties to the system, unlike the structure investigated in the present study.} To find combinations of the structural parameters which lead to maximal polarization rotation, we perform a greedy optimization with respect to $\theta$, $D$ and $\w$. We demand the output power $P_T$ to stay above a minimum value $P_{\text{min}}$ and the eccentricity $e_T$ to be larger than a minimum value $e_{\text{min}}$. In this way, our search is confined to output plane waves with strong amplitudes (otherwise any manipulation would concern a tiny portion of the input signal) and as-linear-as-possible polarization (since one cannot define meaningfully polarization tilt in quasi-circularly polarized waves). Within this set of transmission signals, we select those with high tilt angles $|\Delta\theta|$.

In Fig. \ref{fig:Figs3}(e), we show the loci of the moving tip of $\Einc$ and $\Etra$ vectors with time, for a large tilt $|\Delta\theta|=31^o$ concerning the $P_T=75\%$ of the incident power. The polarization of the transmitted field is elliptical as expected, but the eccentricity $e_T$ is high. Tilts of similar or even smaller angles have been reported in magnetically-biased devices \cite{GiantFaradayRotation, HugeFaradayRotation}. Note that if a biaxially anisotropic structure is illuminated normally with a plane wave whose electric field is parallel to one of its two axes ($\theta=0^o,~ 90^o$), the output  (reflected or transmitted) would have exactly the same polarization and thus $\Delta\theta=0$; therefore, a sweep with respect to angle $\theta$ is essential.

In the same spirit, we can optimize our simple structure in order to create a polarization transformer from the linear incidence to an almost circular transmission. Again, we keep the power carried by $\Etra$ above a desired threshold,  $P_T>P_{\text{min}}$, to avoid substantial reflections. On the contrary, the eccentricity of the polarization ellipse of transmitted field is now kept below a maximum value,  $e_T<e_{\text{max}}$, since we aim at less oblate ellipses. In Fig. \ref{fig:Figs3}(f), we show the input and output of a structure which achieves a relatively good conversion from a linearly polarized incident signal to a transmitted field with quasi-circular transmission concerning the $P_T=58\%$ of the incident power. The same design can work inversely, namely as a circular-to-linear converter, since the structure is reciprocal (no external magnetic bias).

In conclusion, we have provided evidence of how a slab of BP can be used  effectively to manipulate the properties of incident polarization. The transmitted portion can have polarization significantly and controllably different than the incident component. This level of control of the polarization states of light is an important utility to be exploited in the modeling and design of ultra-efficient components and modules, in the context of a  wide range of state-of-the-art photonics applications.

We acknowledge support by the School of Science and Technology of Nazarbayev University (C.A.V.); EFRI 2-DARE NSF Grant No. 1542807 (M.M.); ARO MURI Award No. W911NF14-0247 (E.K.); the E.U. program  H2020-MSCA-RISE-2015-691209-NHQWAVE (M.M.). We used computational resources on the Odyssey cluster of the FAS Research Computing Group at Harvard University.


\end{document}